\documentclass[aps,prd,twocolumn,showpacs,groupedaddress]{revtex4}
\usepackage{epsfig}
\usepackage{longtable}
\usepackage{amsfonts}

\newcommand{\be}{\begin{equation}}
\newcommand{\ee}{\end{equation}}
\begin{document}

\title{A new photon recoil experiment: towards a determination of the fine structure
constant}
\author{Holger M\"uller, Sheng-wey Chiow, Quan Long, Christoph Vo, and Steven Chu}
\affiliation{Physics Department, Varian Bldg., Room 226, Stanford University, Stanford, CA 94305-4060; Phone: (650) 725-2354, Fax: (650) 723-9173}
\email{holgerm@stanford.edu.}
\date\today
\begin{abstract}
We report on progress towards a measurement of the fine
structure constant $\alpha$ to an accuracy of $5\times 10^{-10}$ or better by measuring the ratio $h/m_{\rm Cs}$ of the Planck constant $h$ to the mass of the cesium atom $m_{\rm Cs}$. 
Compared to similar experiments, ours is improved in three significant ways: (i)
simultaneous conjugate interferometers, (ii)
multi-photon Bragg diffraction between same internal states, and
(iii) an about 1000 fold reduction of laser phase noise to -138\,dBc/Hz. Combining
that with a new method to simultaneously stabilize the phases of
four frequencies, we achieve 0.2\,mrad
effective phase noise at the location of the atoms. In
addition, we use active stabilization to suppress systematic effects due
to beam misalignment. 

\end{abstract}


\maketitle


\section{Introduction}
The fine structure constant $\alpha$ \cite{Sommerfeld} describes the strength of the
electromagnetic interaction and is thus important in all of
physics, from elementary particle, nuclear and atomic physics to
mesoscopic and macroscopic systems; an improved measurement will
thus be a basis for many applications in both fundamental and
applied science. This also implies that $\alpha$ can be measured by a wide variety of
precision experiments, see Table
\ref{alphavalues} \cite{CODATA}. This also provides one of the
best measures of the overall consistency of our basic theories and
experimental methods. Measurements of $\alpha$ have been made using, for example, the quantum Hall effect,
the ac Josephson effect, or neutron interferometry. (The latter value also depends on the electron and neutron masses and crystal lattice spacings.) It is unclear whether the uncertainties of these methods can dramatically be reduced in the near future.

\begin{table}[b]
\centering \caption{\label{alphavalues} Precision measurements of
the fine-structure constant \cite{CODATA}. See Table XV in Ref. \cite{CODATA} for a more complete overview.}
\begin{tabular}{cccc}\hline
Description & $\alpha^{-1}-137.03$ & uncertainty \\ & $10^{-3}$ & ppb \\ \hline
Electron
g-factor \cite{VanDyck,VanDyck2} & 5.998 80(52) & 3.8 \\ Photon Recoil $h/m_{\rm Rb}$ \cite{Biraben}$^1$ & 5.998 78(91) & 6.7 \\ Photon Recoil $h/m_{\rm Cs}$ \cite{Wicht,Gerginov06} & 6.0001(11) & 7.7 \\ 
Quantum Hall effect $R_K$ \cite{Jefferey,Jefferey2} & 6.0037(33) & 24 \\ Neutron
Interferometry \cite{Krueger} & 6.0015(47) & 34 \\ AC Josephson effect
$\Gamma'_{p-90}(lo)$ \cite{Williams} & 5.9880(51) & 37 \\ Muonium Hyperfine
structure \cite{Liu} & 5.9997(84) & 61
\\ \hline CODATA 2002 adjusted value & 5.999 11(46) & 3.3 \\ \hline
\end{tabular}\\
{\footnotesize $^1$This recent measurement is not part of the CODATA2002 adjustment.}
\end{table}

\begin{figure}
\centering
\epsfig{file=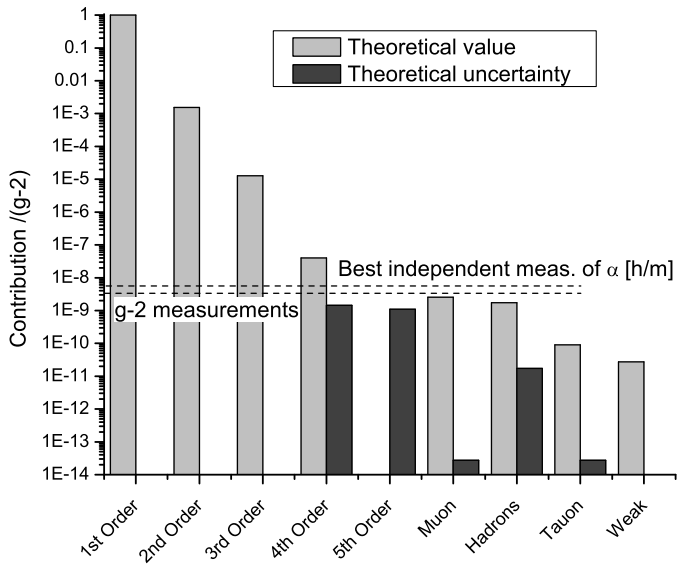, width=0.5\textwidth}
\caption{\label{g-2} Relative magnitude of the terms entering the QED calculation of $g_e-2$ together with the present accuracies of relevant experiments.}
\end{figure}

The best precision measurements of $\alpha$ to date, however, are based on the following two different principles: From a 
measurement of the electron's anomalous magnetic moment $g_e-2$, $\alpha$ can be calculated using quantum electrodynamics. This calculation \cite{Kinoshita} is probably the most accurate prediction in science (that is not just a simple combination of integers or mathematical constants such as $e$ and $\pi$). It includes electron terms as well as contributions from the muon, the tauon, hadrons, and even the weak interaction, see Fig. \ref{g-2}. At the current experimental accuracy, however, only the electron terms (up to the 4th order) are relevant. From the best current data on $g_e-2$, $\alpha$ is obtained to a precision of 3.8\,ppb. On the other hand, $\alpha$ can be obtained from measurements of $h/m$, the ratio of the Planck constant to the mass of an atom. The two best experiments both lead to a similar accuracy of about 7\,ppb in $\alpha$: Biraben and co-workers \cite{Biraben} used Rubidium atoms and Wicht {\em et al.} (the predecessor of the experiment described here)\cite{Wicht,Wichtproc} Cesium. The accuracy of the latter result may improve by up to a factor of two when an accurate theoretical estimate of the largest systematic effect has been made, the index of refraction of Cs atoms. This method does not rely on involved QED calculations. Thus, on the one hand, it is not limited by the accuracy of present QED theory. On the other hand, a comparison to $\alpha$ as obtained from $g_e-2$ amounts to a test of QED. The present values of $\alpha$, obtained from both $g-2$ and $h/m$ experiments, agree within the overall experimental error of about 8\,ppb. This corresponds to one of our best tests of QED and is an impressive accomplishment of modern physics. 

Recently, Gabrielse {\em et al.} have repeated the $g_e-2$ measurement and anticipate a
precision in $\alpha$ of $\sim 0.6-1$\,ppb \cite{Odom}. This would open up the possibility to increase the accuracy to which QED is experimentally tested by more than a factor of ten by an improved measurement of $h/m$. At this level of precision, the influences of the vacuum polarization caused by muons (2.5\,ppb) and hadrons (1.7\,ppb) shown in Fig. \ref{g-2} would become visible. Thus, we are not only looking for the tenth digit of $\alpha$, but also for the first digit of these effects. Many new discoveries in physics have come out of experiments that have pushed the boundaries of applicability of
our physical theories.

A preliminary measurement of the anomaly $g_\mu-2$ of the gyromagnetic ratio of the muon to 1\,ppm (part per million) at the Brookhaven laboratory \cite{Bennett,Hertzog} makes this particularly exciting: Data obtained over five years deviate from the QED prediction by 0.9-2.4 standard deviations (equivalent to 0.5-1.2\,ppm); the exact deviation depends on the method employed to estimate the hadronic vacuum polarization. If other explanations would fail, however, the discrepancy could be a first indication for supersymmetric particles. Our experiment with its sub-ppb accuracy will allow to resolve the hadronic influence for the electron and can thus contribute to a
clarification of these issues. 

\subsection*{Outline of the paper}

This paper describes how we want to reach the above accuracy of 0.5\,ppb (or better) in $\alpha$. Since the phase of the atomic wave function is measured against the phase of laser beams, the accuracy is based on a laser system of high phase stability. While many other systematic effects also need to be taken into account, in this paper we focus on the construction of the laser system. Section \ref{basicidea} describes the basic principle of the measurement of $h/m$; the atom interferometric realization is then described in Sec. \ref{atominterf}. The main part of the paper, Sec. \ref{locks}, describes the feedback loops that we use to obtain a drastic reduction of laser phase noise, the most important noise contribution in most atom interferometers. We concentrate on the role and the interaction of the secondary phase locks: These take out the noise picked up by the light on its way from the lasers to the atoms and thus reduce the noise as seen by the atoms. We believe that these technical aspects are interesting in the context of present and future atom interferometry projects (such as spaceborne experiments).

\section{Basic idea}\label{basicidea}

The principle of our experiment is to measure the recoil frequency
\be
f_r=\frac{hf^2}{2mc^2}
\ee
of an atom with the mass $m$ that
absorbs or emits a photon having the frequency $f$, where $h$ is
the Planck constant and $c$ the velocity of light: If an atom
absorbs a photon, it also absorbs the photon momentum $p=hf/c$ and
thus acquires a kinetic energy of $p^2/(2m)=h^2f^2/(2mc^2)$. Thus,
transitions between the ground and the excited states can take
place if the light frequency is
\begin{equation}
f=f_{eg}\pm f_r\,, 
\end{equation} 
where $hf_{eg}$ is the energy of the excited state of the atom relative
to the ground state in the rest frame of the atom. The positive
sign yields absorption, the negative stimulated emission. In our
experiment, we use the D2 transition in cesium atoms, so $m=m_{\rm
Cs}$ is the mass of the cesium atom and $f\simeq f_{D2} \simeq 352\,$THz. Thus,
$f_r \simeq 2.066\,$kHz. A very basic experiment to measure $f_r$ would measure the difference in the frequencies of light that produce stimulated emission and absorption in Cs atoms.

By using the relation
\be\label{alphaequation}
\alpha^2=4R_\infty\frac{f_rc}{f_{D2}^2}\frac{m_{\rm Cs}}{m_u}\frac{m_u}{m_e}\,,
\ee
we can express $\alpha$ by the recoil frequency, the Rydberg constant $R_\infty$
and ratios between $m_{\rm Cs}$, the atomic mass unit $m_u$, and the electron mass $m_e$. This can easily be verified by inserting the definitions $R_\infty=m_e e^4/(8c\epsilon_0^2h^3)$ and $\alpha=e^2/(4\pi\epsilon_0\hbar c)$. The quantities entering this relation are known to high accuracy from frequency measurements (hydrogen spectroscopy
for $R_\infty$ and measuring the ratio of cyclotron frequencies in Penning
traps for the mass ratios), see Tab. \ref{auxiliaryvalues}. The
uncertainties in these quantities have been improved recently by new measurements of $m_e/m_u$ and the Cs transition frequency and combine to  0.24\,ppb in $\alpha$ [note that most uncertainties enter $\alpha$ with half their value due to Eq. (\ref{alphaequation})]. However, for both the mass ratios as well as the Rydberg constant, there is a clear potential for significant improvements in the near future. 
But already with the present uncertainties, we can obtain $\alpha$ to an accuracy of up to about 0.3\,ppb. 

\begin{table}[b]
\centering \caption{\label{auxiliaryvalues} Quantities entering
the determination of $\alpha$ from $h/m_{\rm Cs}$.}
\begin{tabular}{cccc}\hline Quantity & value & uncertainty & Ref. \\ & & ppb & \\ \hline
$m_e/m_u$ & $5.485 799 0945(24)\times 10^{-4}$ & 0.44 & \cite{CODATA} \\
$m_{\rm Cs}/m_u$ & 132.905 451 931(27) & 0.20 &
\cite{Bradley} \\ 
$f_{D2}$ & 351 725 718.4744(51)\,MHz & 0.015 &
\cite{Gerginov} \\ $R_\infty$ &
10 973 731.568 525(73)m$^{-1}$ & 0.0066 & \cite{CODATA} \\
\hline
\end{tabular}
\end{table}

\section{Atom interferometry for measuring $h/m$}\label{atominterf}
\subsection{Previous experiment at Stanford}

\begin{figure}
\centering \epsfig{file=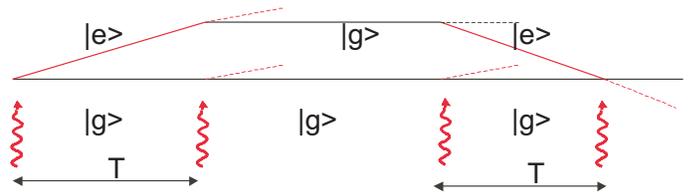,width=0.5\textwidth}
\caption{\label{interferometer} Basic (Ramsey-Borde) atom
interferometer with four $\pi/2$ laser pulses as beam splitter.}
\end{figure}

The measurement of $h/m$ is extremely challenging: The recoil frequency shift of about 2\,kHz out of $f_{\rm D2}\simeq 352\,$THz is minuscule --- it is even small compared to the natural linewidth
$\Gamma\sim2\pi\times5\,$MHz of the transition. Therefore, several refinements were introduced in the previous experiment:
\begin{itemize}
\item Two-photon transitions between two hyperfine ground states
are made. The two ground states are long-lived and thus their
linewidth is no longer limited by the lifetime of the excited
state. Besides, the two photon momenta increase the recoil shift
to $4f_r\sim 8$\,kHz, as the effective wavevector $\mathbf k_{\rm eff}=\mathbf k_1-\mathbf k_2$ is the (vector) difference of the wavevectors $\mathbf k_{1,2}$ of the counterpropagating beams.
\item Atom interferometry [see Fig.
\ref{interferometer}] can dramatically improve the
resolution. The beam splitters are
formed by $\pi/2$ laser pulses that transfer the atoms from a
state $\left|g, \mathbf p_0\right>$ (where $g,e$ denote the ground and excited states and $\mathbf p_0$ the initial external momentum) to a state $\left|e, \mathbf p=\mathbf p_0+\hbar
\mathbf k_{\rm eff}\right>$ with 50\% probability. By interfering two
atomic paths that have experienced the recoil $\hbar \mathbf k_{\rm eff}$
at different times separated by $T$, one obtains interference
fringes with a width of $1/T$. The atom source is an atomic fountain in which atoms cooled to near the recoil temperature $T_{\rm rec}$ (200\,nK for cesium) \cite{Treutlein} are launched and are under free-fall conditions for up to about 1\,s. Fringe widths of 4\,Hz are 
practical, an improvement of 6 orders of magnitude compared to the
basic experiment (in which the linewidth is either limited by the natural linewidth of 5\,MHz or even the Doppler linewidth of the thermal atoms of $\sim 300\,$MHz).
\item We compare the measurements of two
``conjugate" interferometers, i.e., ones where the recoil points
vertically upwards and downwards, see Fig. \ref{interferometer2}. Therefore, the signal is doubled while the fringe shift
due to gravity cancels out (as do numerous other systematic
effects). \item By inserting up to $N=30$ additional excitations
into the interferometer paths, we increase the recoil to $4f_r(N+1)$.
\end{itemize}
This describes the final stage of that experiment, which was improved several times. For simplicity, we left out several details like the use of adiabatic transfer to perform the beam splitters and $\pi$-pulses, that will be found in the literature \cite{WeissPRL,Weiss,Hensleyvib,Hensley,Wicht,Wichtproc}. 

\subsection{New experiment}

\begin{figure}
\centering
\epsfig{file=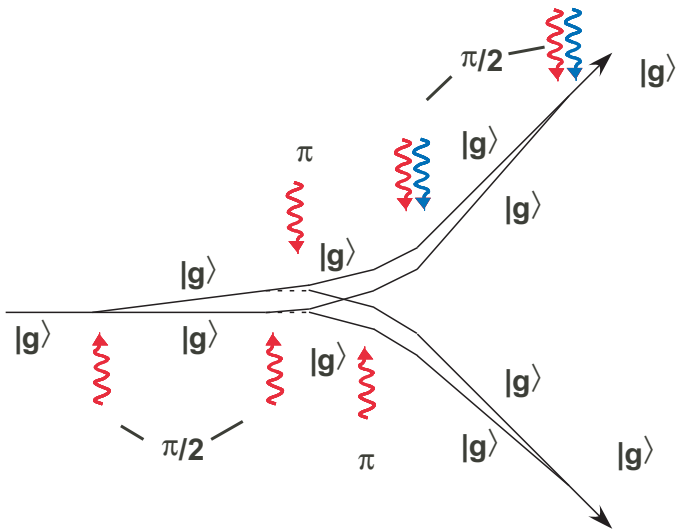,width=0.5\textwidth}
\caption{\label{interferometer2} SCIs used in this experiment. The
last two $\pi/2$ pulses contain four frequencies in two pairs, to
simultaneously address both conjugate interferometers.}
\end{figure}

\begin{figure}
\centering
\epsfig{file=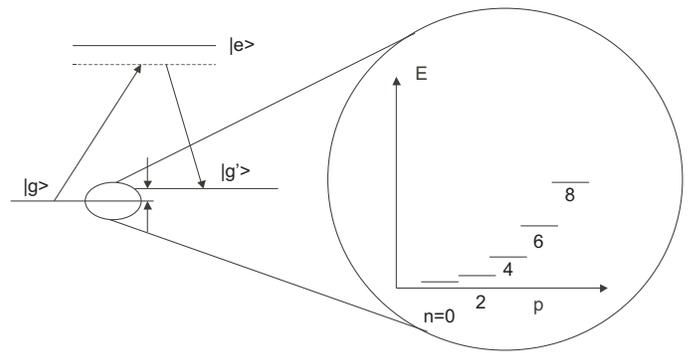,width=0.5\textwidth}
\caption{\label{Bragg} Bragg diffraction: A Raman transition (left) transfers the atom into another ground state. In Bragg diffraction, a $n-$photon transition between the ground state and the same ground state is performed. However, this applies $n$ times the recoil momentum $\hbar k$ to the external motion of the atom and thus transfers it into a momentum energy level at $n^2$ times the recoil energy, transferring the atom to a higher level of (external) kinetic energy.}
\end{figure}

In this section, we will describes the major
improvements that are made in the new experiment to achieve better accuracy:
\begin{itemize}
\item Simultaneous conjugate interferometers (SCIs): The use of adiabatic transfer for the beam splitters (as well as the technical difficulty of addressing interferometers simultaneously, that we will deal with in this paper) made it necessary to run the two conjugate interferometers subsequently. This means, however, that time-dependent
disturbances such as vibrations do not cancel between the interferometers. Using SCIs, however, (Fig. \ref{interferometer2}) we can achieve
cancellation for both time-independent and -dependent errors. Obviously, however, if the vibrations enter the phase relationship between the laser beams used in the SCIs, this method is of little use. The major challenge thus is to generate these four laser beams with an accurate phase relationship.
\item Multi-photon transitions between same ($F=3$ or $F=4$, $m_F=0$)
internal states: We replace the transitions between hyperfine
states by multi- ($n$-) photon transitions (``Bragg diffraction")
$\left|g, p=p_0\right>\rightarrow\left|g, p=p_0+n\hbar k_{\rm
eff}\right>$ between different momentum states, but within a
single internal state of the atom (Fig. \ref{Bragg}). This strongly reduces the influence of magnetic and
electric fields, since the internal state is never changed: Thus, the sensitivity to external fields is now the same in all interferometer paths and thus cancels. $n-$photon transitions also increase the
momentum transferred to the atoms. The phase of the atomic wave
function $\Psi$ is proportional to the momentum squared, and thus
to $n^2$. The signal to noise ratio can thus be increased by $n^2$. Bragg diffraction of atoms has been demonstrated in atomic beam interferometers \cite{Giltner,Giltner2}. Beam devices, however, have shorter interaction times and lower signal to noise ratio than the atomic fountain interferometers used for precision measurements. 
\item Strong laser noise reduction: The most severe
limitation of the previous experiment was laser phase noise, arising, for example, from imperfections of the laser phase stabilization, mechanical vibrations of optics and fibers, and air currents. It took about a
week of integration time to obtain the final resolution of a few
ppb. Experience has shown that the resolution obtained in a few
days corresponds to the accuracy possible in one year, since the
experiment has to be repeated many times in order to check for
systematic effects. Therefore, we reduce the laser noise about
1000 fold to hopefully achieve a ppb level resolution in minutes rather than
weeks. This reduction is achieved by an extremely tight phase-lock, common-mode rejection between the conjugate interferometers, and secondary phase stabilization loops that measure the phase near the location of the atoms. More details on this will be found in Sec. \ref{locks}.
\end{itemize}

\subsection{Phase of the SCIs}

The phase of the SCIs shown in Fig. \ref{interferometer2} is given as follows: We use the usual Wentzel-Kramers-Brillouin method. There are four $\pi/2$ pulses for each atom group. The Hamiltonians describing the interactions of the atoms with these laser pulses are proportional to $\exp(i n\phi_{1-4}/2)$, where $\phi_{1-4}$ are the phases of the pulses. The factor of $n/2$ arises because an $n-$photon transition corresponds to an $n/2$th order Bragg diffraction. Thus, the transition amplitudes are proportional to the interaction to the power of $n/2$. We add an accent to denote the phases for the second interferometer. The phases of the two interferometers are denoted $\Phi_{1,2}$. We obtain
\begin{eqnarray}
\Phi_1&=&2\omega_rn^2(N+1)T+\frac n2 (-\phi_1+\phi_2+\phi_3-\phi_4)\,,\nonumber \\
\Phi_2&=&-2\omega_rn^2(N+1)T+\frac n2 (-\phi'_1+\phi'_2+\phi'_3-\phi_4)\,.\nonumber \\
\end{eqnarray}
In addition, there is a phase shift due to gravity, which we did not include as it drops out of the final result. For the first two pulses, there are only two velocity groups (Fig. \ref{interferometer2}), so we can take $\phi_{1,2}=\phi'_{1,2}$. The phase difference of the individual interferometers is
\begin{eqnarray}\label{totalphase}
\Phi_1-\Phi_2&=&4\omega_rn^2(N+1)T+\frac n2(\phi_3-\phi'_3-\phi_4+\phi'_4)\nonumber \\ &\equiv & 4\omega_rn^2(N+1)T+n\phi \,.
\end{eqnarray}


\section{Interferometer laser system and noise reduction}\label{locks}

At the heart of the experiment is a laser system generating the
light pulses for the beam splitters in the interferometer. The accuracy
that we aim for requires a phase noise much lower than
demonstrated in previous experiments; and the SCIs require four simultaneous,
phase locked frequencies. The system has to supply these
frequencies in Gaussian pulses of the appropriate length for
obtaining $\pi/2$ and $\pi$ pulses. 

\subsection{Requirements}

To simultaneously address the two interferometers, four laser frequencies $f_{1-4}$ are needed, which have to satisfy resonance conditions in the frame of the falling atoms. Technically, the phase $\phi$ is given by phases of four simultaneous laser beams. The first interferometer will be addressed by the frequencies $f_1$ and $f_4$. The second interferometer is addressed by $f_2$ and $f_3$. Thus, we find 
\begin{equation}
\phi=2\pi[f_1-f_4-(f_2-f_3)]T\equiv 4\pi f_mT
\end{equation} 
for the laser phase entering the interferometers. The measurement process for $h/m$ will be scanning $f_m$ and measuring the interference fringes in the two individual atom interferometers. In the center of the fringes, we have from Eq. (\ref{totalphase}),
\begin{equation}
2f_r n (N+1)+ f_m =0\,.
\end{equation}

We can express the relevant difference frequencies by a common-mode frequency $f_0$ and a differential-mode frequency $f_m$:
\begin{eqnarray}
f_1-f_4&\equiv &f_0+f_m\,,\nonumber \\ 
f_2-f_3&\equiv &f_0-f_m\,. 
\end{eqnarray}
The common-mode frequency $f_0$ accounts for the Doppler effect due to the gravitational motion of the atoms. In the laboratory frame, $f_0$ is therefore swept at a rate of $2gf_{D2}/c \simeq 23$\,MHz/s (the factor of 2 is because one of the counterpropagating beams is seen blue- the other red-shifted by the atoms, respectively). While each of $f_{1-4}$ contain the large gravitational offset $f_0$ and the influence of vibrations, the combination $f_m$ is independent of these. Since the combination $(f_1-f_4)-(f_2-f_3)=2f_m$ provides the reference for the measurement of $f_r$ and thus $h/m$, the frequency generation scheme has to generate this combination with the lowest possible noise. The averages $(f_1+f_4)/2=(f_2+f_3)/2$ are approximately 10\,GHz blue detuned relative to the F=3 to F=5 transition of the D2 line at 852\,nm.

For a $10^{-10}$ accuracy in $4n(N+1)f_r\sim 0.88\,$MHz (where we assumed $n=N=10$), we need to
know the relevant combination of frequencies to 0.1\,mHz, or
$3\times 10^{-19}$ of the laser frequencies. More specifically, if
the overall phase uncertainty in $f_m$ amounts to $\delta \phi$,
then the relative error in $f_r$ in the differential measurement
between conjugate interferometers would be 

\begin{equation}\label{phasenoise}
\frac{\delta f_r}{f_r}=\frac{\delta \phi}{4\pi
n(N+1)f_rT}\,,
\end{equation}
where $T\sim 0.2\,$s. An effective phase noise of
20\,mrad, for example, with $N=10, n=10,$ and T=0.2\,s, would give an error of 36\,ppb in $h/m$. If this is reached for one individual launch (every two seconds), it would be 1\,ppb in $h/m_{\rm Cs}$ and 0.5\,ppb in
$\alpha$ within a few hours of integration.

The frequency $f_0$, on the other hand, drops out of the final result. However, a large noise in $f_0$ would make it impossible to measure interference fringes within the individual interferometers. By a simultaneous analysis of the combined data using an ellipse fit \cite{Markfit}, it is still possible to measure the differential signal. However, for different applications and as a tool for characterizing systematic effects, one wants to have fringes from the individual interferometers. Therefore, we need a good phase noise performance in $f_0$ as well. 

\subsection{Basic setup}

\begin{figure*}
\centering \epsfig{file=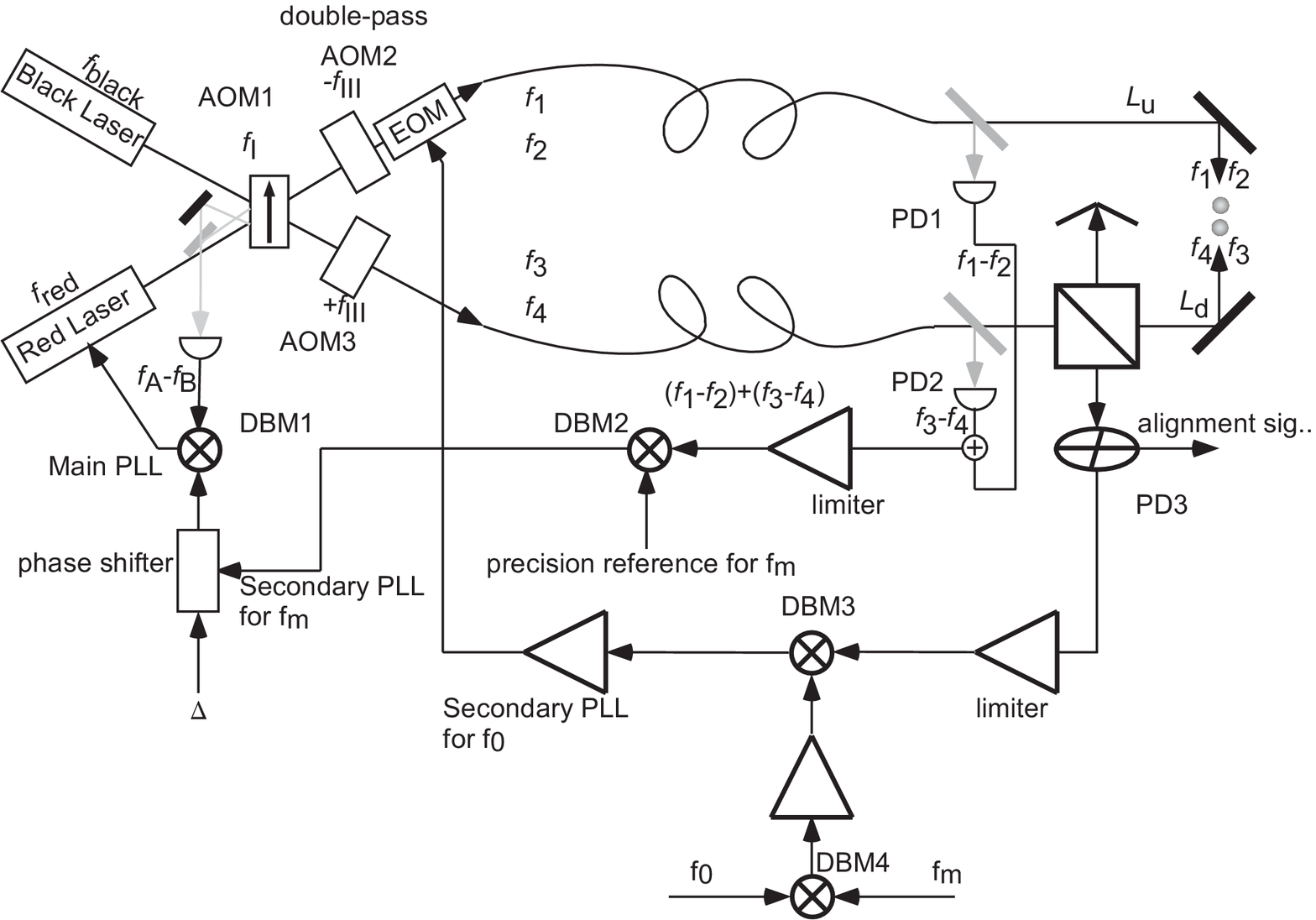, width=0.8\textwidth}
\caption{\label{conjugate} Frequency generation and phase locking scheme for SCIs.}
\end{figure*}

Two frequencies are generated from two Coherent 899 Ti:sapphire lasers that will be denoted red and black in this paper (Fig. \ref{conjugate}). Compared to diode lasers \cite{Wieman}, Ti:sapphire lasers provide higher output power, which is important for high-order Bragg diffraction with high efficiency. The black laser ($\sim 1.4$\,W output power) is pumped by a Coherent Verdi-10. The red laser ($\sim 1.6\,$W) contains an intra-cavity electro-optic modulator (EOM) for phase locking. Because of the losses associated with the insertion of the EOM it needs a higher pump power, which is provided by a Coherent Innova argon-ion laser that is running at 15.6\,W output power. (Without the EOM, the power of the red laser can be 2.0 W at this pump power.) The light of both lasers is brought into the setup by two single-mode fibers (not shown), to decouple the alignment of the setup and the one of the lasers. The fibers provide about 1.0 and 1.2\,W at their outputs.

The frequency of the black laser is stabilized to a frequency which is 10\,GHz higher than the one of the $F=3\rightarrow F=5$ D2 manifold. This is done by phase-locking it at 10\,GHz offset (given by a microwave synthesizer) to a diode laser that is itself frequency stabilized to a resonance in a cesium vapor cell. 

\begin{figure}
\centering \epsfig{file=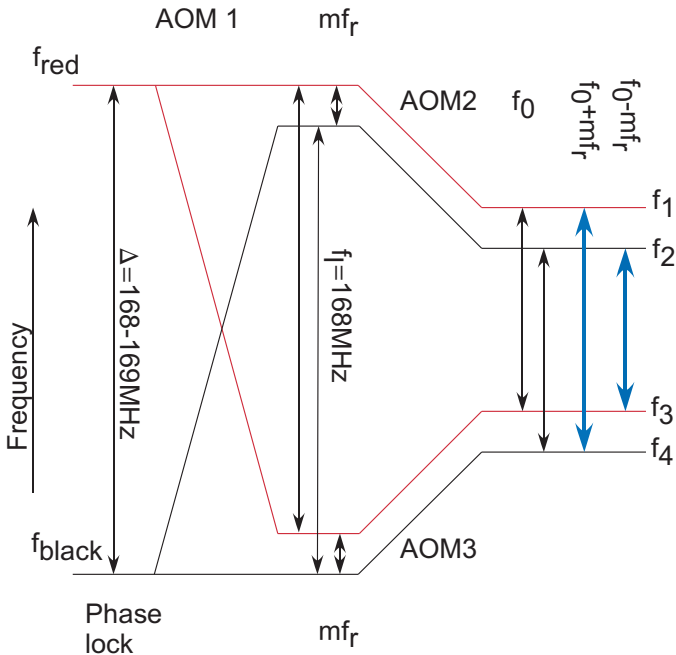, width=0.5\textwidth}
\caption{\label{freqgen} Frequency generation. $\Delta$ is made
variable in order to set the reference for $f_m=\Delta-f_{\rm I}$
between 0 and about 1\,MHz. The frequencies of AOM1 and AOM2 are
swept to account for the time dependent Doppler effect due to the
free-fall motion of the atoms. $f_{\rm I}$ is fixed.}
\end{figure}

The frequency difference $f_{\rm red}-f_{\rm black}\equiv \Delta$ of the lasers is phase locked to a variable frequency around 168\,MHz (Fig. \ref{freqgen}) by the ``primary" phase lock as described in Ref. \cite{PLL}. It uses the intracavity EOM as a feedback path with 10\,MHz closed-loop bandwidth, achieved using broadband radio-frequency amplifiers to drive the EOM and a passive network to set
the frequency response. The speed is actually limited by the
delay due to the $\sim 5$\,m signal path between the EOM and the
detector. 

\begin{figure}
\centering \epsfig{file=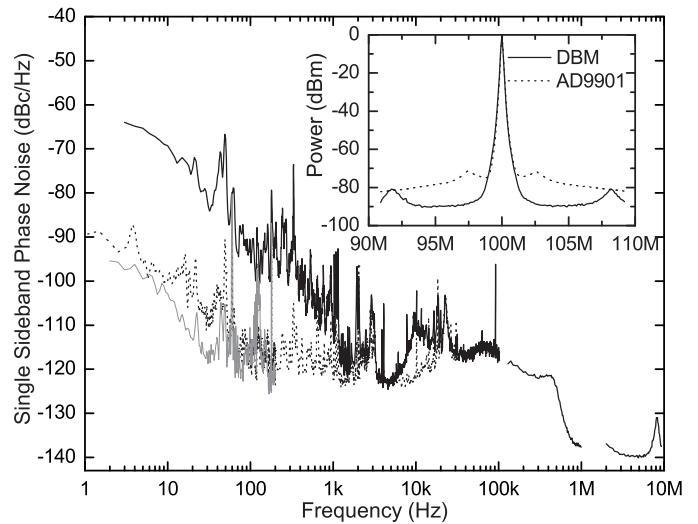, width=0.5\textwidth}
\caption{\label{PLLphasenoise} Phase noise spectrum for
$2f_m$; the phase noise of $f_m$ is a factor of two lower. Solid curve, phase lock before fiber
only; dashed curve: phase lock after fiber; gray curve, purely
electronic phase noise measured with a separate phase detector driven by the same signals as DBM1 (Fig. \ref{conjugate}). The gaps at 100 kHz and 1 MHz are due to different
range settings between data sets. Inset, phase noise spectra
of analog (DBM) and digital phase detectors (AD9901). Resolution bandwidth,
100 kHz.}
\end{figure}

We achieve a phase-noise of -138\,dBc/Hz (i.e., the noise power spectral density expressed in dB within one Hz bandwidth, referred to the carrier) measured 1\,MHz from the carrier (Fig. \ref{PLLphasenoise}, solid graph), perhaps the lowest phase noise achieved in any phase locked laser system. The phase noise was measured by taking a beat note between the lasers using a separate photodetector with a HP 8590B spectrum analyzer. The low frequency part below 100 kHz was measured from that signal with an SR 785 fast Fourier transform analyzer. See Ref. \cite{PLL} for details. However, even the best phase-lock does not guarantee that the light as seen by the atoms has low (or even acceptable) noise, because the transmission of light through fibers \cite{fiberpaper}, the vibration of optical elements, and even the fluctuating index of refraction of air due to air currents seriously degrade the phase stability. Therefore, we use an elaborate system of feedback loops to cancel the phase noise as seen by the atoms. In this paper, we will concentrate on these extra feedback loops and their interrelation.

Although it has been pointed out in the literature (e.g., \cite{Wieman}), we note here that there are many common mistakes in frequency and phase-noise measurements. The most common one is to infer the noise from the error signal in a closed phase or frequency locked loop (``in-loop measurement"). There is no certainty that all noise components are revealed in such a signal. For example, if the beam splitter in the beat setup of the primary phase lock (Fig. \ref{conjugate}) vibrates, this causes the phase of the beat to be unstable. However, this instability is removed by the lock, with the result that its negative image now appears in the red laser light. For this and other reasons (such as imperfections of the phase detector), the in-loop measurement often underestimates the real noise, by as much as 40\,dB in some cases. A conservative method to measure phase noise is to overlap the beams on another beamsplitter elsewhere in the setup and use a separate beat detector (``out-of-loop signal"). All phase noises quoted in this work have been measured using this method.

The beams of the red and the black lasers are overlapped on an acousto-optic modulator (AOM1, Fig. \ref{conjugate}) that is running at a frequency $f_{\rm I}\simeq 168$\,MHz at 50\%
deflection efficiency. Two pairs of beams are thus obtained, each consisting of an un-deflected beam from one laser and a deflected and frequency shifted beam from the other laser. Each of them has a power of about 500\,mW. The frequencies of them are
\begin{eqnarray}
{\rm first \, pair} \left\{\begin{array}{c} f_{\rm red}\\ f_{\rm black}+f_{\rm I} \end{array} \right. \\
{\rm second \, pair} \left\{\begin{array}{c} f_{\rm black} \\ f_{\rm red}-f_{\rm I} \end{array} \right.
\end{eqnarray}
(Fig. \ref{freqgen}). Within each pair, the frequency difference is thus $\Delta-f_{\rm I}$. This will be the reference for measuring the recoil shift. Each pair then gets frequency shifted by the double-pass AOMs 2 and 3 by $f_{\rm III}$. After that, the frequencies are
\begin{eqnarray}
f_1&=&f_{\rm red}-f_{\rm III}\,, \nonumber \\ f_2&=&f_{\rm black}+f_{\rm I}-f_{\rm III}\,,\nonumber \\
f_3&=&f_{\rm red}-f_{\rm I}+f_{\rm III}\,,\nonumber \\ f_4&=&f_{\rm black}+f_{\rm III} \,.
\end{eqnarray} 
These AOMs are also used to amplitude-modulate the beams with a Gaussian envelope that is proportional to $\exp[-t^2/(2\sigma^2)]$ with a $\sigma$ of the order of 20$\,\mu$s. The AOMs need to be located before the fibers (Fig. \ref{conjugate}). Located after the fibers, they could impair the pointing stability and the shape of the beams that address the atoms. Pointing stability and a high quality of the spatial mode pattern are important, because the wave-fronts of the beams define the ``ruler" that measures the phase of the atomic states.  

Active feedback is used to control the envelope waveform to obtain an accurate Gaussian envelope. Although just two actuators are provided to stabilize the power of four beams, the amplitude stabilization is arranged such that it keeps the Rabi frequency of the transitions in each of the SCIs constant. However, we will focus on the generation of frequencies here and thus skip the description of this part. 

Each pair then enters its own single-mode fiber with the same efficiency for each of the frequencies. With a well-aligned setup, we achieve about $180\,$mW of output power per frequency, i.e., $600\,$mW total. For the frequency $f_m$ that is the reference for the recoil frequency, we now have 
\begin{equation}
2f_m=f_1-f_4-(f_2-f_3)=2(\Delta-f_{\rm I})\,.
\end{equation}

\subsection{Common-mode rejection}

Because of the nearly perfect overlap of the beams within each pair (verified by the similarity of the fiber-coupling efficiencies), any vibrations affect both frequencies nearly equally and thus should have a very slight effect on the measurement of $mf_r$. 

To make the best use of this common-mode rejection, the beat signal used for the primary phase lock are the parasitic reflections from the input surface of AOM1 (Fig. \ref{conjugate}): Any vibrations that cause phase noise before AOM1 are removed by the primary phase lock. Vibrations after AOM1, however, are common mode and thus rejected. Vibrations in the signal path of the beat system after the beam splitter are common to both frequencies and are rejected by controlling the red laser's frequency. The mirror and beam splitter of the beat setup are mounted rigidly using half-inch optics on a common copper block. This reduces the influence of vibrations. The copper block is electronically temperature stabilized to keep the optical path length constant to below a wavelength. As a result, the DC phase between the beat frequencies of the signals in each of the two fibers is long-term stable to much below a wavelength. This will be important for the operation of the secondary phase locks.

\subsection{Secondary phase lock for $f_m$}

There are two significant frequency differences between the laser beams, $f_0$ and $f_m$. The stability of both is improved by two secondary feedback loops that measure the noise after the fibers. Since the one for $f_0$ cannot operate without the other one, we will explain the $f_m$ post-fiber phase lock first. 

The transmission of the signals through fibers and the vibration of optical elements can introduce many radians of phase uncertainty. However, as explained, most of that cancels in our setup. The effective phase noise that is relevant for the interferometer is given by integrating the noise spectral density shown in Fig. \ref{PLLphasenoise}, solid graph, from 1 Hz to about 10\,kHz (these limits are rough, but conservative, estimates based on a pulse duration of tens of $\mu$s and a pulse separation of 0.2\,s). We achieve a $\sim 5$\,mrad rms effective phase noise, already somewhat better than the above aim of
20\,mrad. Still, however, Fig. \ref{PLLphasenoise} (solid graph) shows a pronounced increase of the noise at frequencies below about 10\,kHz. Because of the common-mode rejection inherent in our
setup, this residual low-frequency noise is too large to be explained simply by vibrations, even taking into account the dispersion due to the slight frequency differences between $f_{1-4}$. A clue to the origin of the noise is that the noise at the outputs of both fibers are uncorrelated. An explanation for this noise is a vibration of the beam pointing
at the entrance of AOM1: Since the acoustic wavelength $\lambda_{\rm I}$
of $f_{\rm I}$ within the AOM is only a few microns, a small displacement
$\delta$ makes the laser beams see the $f_{\rm I}$ at phase that differs
by $2\pi\delta/\lambda_{\rm I}$. Since vibrationally induced dilations of the lasers
are uncorrelated, the common mode rejection does not work for this
effect.

To remove these influences, the light is detected on the platform by three photodetectors, PD1-PD3. PD3 is actually a quadrant photodetector, where the sum of the signals from all quadrants is used here; the individual signals are used for detecting and correcting for any misalignment in the counterpropagation of the laser beams, see Ref. \cite{BPM}. For here, we assume that the beams are aligned, so that PD3 can be viewed as a simple photodetector. 

PD1 and 2 detect the output of each fiber. Their photocurrents $I_{1,2}$ have a Gaussian envelope with a width of $\sigma$ and oscillate at $f_m$. The signals are proportional to  
\begin{eqnarray}
I_1&=&e^{-t^2/(2\sigma^2)}\sin^2[\pi (f_1-f_2) t] \,\nonumber\\
I_2&=&e^{-t^2/(2\sigma^2)}\cos^2[\pi (f_3-f_4) t]\,.
\end{eqnarray}
For now, we assume a perfect Gaussian envelope. For the purpose of this discussion, we assume that $f_{1-4}$ are defined at the locations of the beat detectors. For the secondary phase lock, a phase detector is used that basically measures the timing of the zero crossings of a signal. Therefore, taking the difference generates a signal that has no DC component
\begin{eqnarray}
I_1-I_2 =-e^{-t^2/(2\sigma^2)}\cos[\pi (f_1-f_2+(f_3-f_4)) t] \nonumber \\ \times \cos[\pi (f_1-f_2-(f_3-f_4)) t]\,.
\end{eqnarray}
Now because of the common mode rejection described above, $f_1-f_2$ is very similar to $f_3-f_4$ (the phase error is of the order of a few mrad) even without the secondary phase-lock, so that we can assume $\cos[\pi (f_1-f_2-(f_3-f_4)) t]\approx 1$. The remaining oscillating term has the frequency $[f_1-f_2+(f_3-f_4)]/2=[f_1-f_4-(f_2-f_3)]/2$, which we want to phase-lock to a reference at $f_m$. The subtraction $I_1-I_2$ is performed by wiring the photodetectors in series and taking off the differential signal at their connection. This ensures that $I_1$ and $-I_2$ experience the same gain and phase shift in the signal preamplifier. 

After signal preamplification, the signal is high pass filtered by a first order filter to reject a residual DC signal due to an imbalance of the average intensities of the light on PD1 and PD2. Technically, the desired Gaussian envelopes can only be generated with a certain inaccuracy. This might result in slightly different envelope waveforms for both detectors. The DC part of this is rejected by the high-pass, whereas the time-dependent part can only partially be rejected. The subsequent electronics basically detects the phase via the time of the zero-crossings of the signal. Any offset due to envelope waveform imbalance thus modulates the detected phase. However, while fluctuations in the imbalance will lead to noise, an imbalance cannot lead to a systematic shift of the measurement of  $f_r$, since on average they will cancel in the readout of the interferometer phase since it is common to all $\pi/2$ pulses (unless the fluctuations are systematically different for the third and the fourth $\pi/2$ pulse in each run.)

The signal is then amplified by a limiting amplifier made of cascaded differential amplifiers as described in Ref. \cite{Hobbs}. This has a very large gain and will be saturated by the signal throughout the pulse duration, although the (truncated) Gaussian envelope has an amplitude ratio of 1000:1 between peak and the lowest value. A square-wave of constant amplitude is thus generated, which is fed to DBM2 (Fig. \ref{conjugate}) used as phase detector. It compares the phase of the beat signal to the one of a reference signal generated by a synthesized function generator with low phase-noise. The detected phase difference is used as the input to a PI controller that adjusts the phase of the reference signal of the primary phase lock. The PI controller is switched off and set to a hold position in the inter-pulse interval.

Because of the common-mode rejection and the temperature stabilized beat setup of the primary phase lock, the secondary phase lock only has to make small corrections that do not vary much between pulses. Therefore, it can be a relatively slow feedback loop of a few kHz bandwidth. 

Without use of this secondary phase lock, $f_m$ is given by $\Delta-f_{\rm I}$. Both $\Delta$ and $f_{\rm I}$ are about 170\,MHz, whereas $f_m$ is of the order of 1\,MHz. Since the phase noise of frequency synthesizers using a reference of a given quality is approximately proportional to the output frequency, it is desirable to take $f_m$ from a low-frequency reference directly rather than as the difference of two higher frequencies. Indeed, the stability of our optical phase locks exceeds the one of a good synthesizer at 168\,MHz, whereas a synthesizer at 1\,MHz can be better. The secondary phase lock allows to do that (Fig. \ref{conjugate}): A reference for $f_m$ is generated directly by a synthesizer, and the secondary phase lock corrects $\Delta$ to keep $f_m$ aligned to that reference. $\Delta-f_{\rm I}$ has to be set to $f_m$ to within a few Hz accuracy by the synthesizer for $\Delta$. Within a typical pulse duration of 200$\,\mu$s, the phase correction necessary is then a few mrad, well within the range of the phase shifter. 

It is very important that noise from vibrational variations of the optical path lengths $\delta L_u,\delta L_d$ from PD1 and 2 to the atoms for the upper and lower beams (Fig.\ \ref{conjugate}) is cancelled out. In the individual paths, it causes optical phase changes of $2\pi \delta L_u \frac{f_{1,2}}{c},2\pi\delta L_d \frac{f_{3,4}}{c}$, which results in phase shifts in the individual atom interferometers proportional to 
\begin{equation}
2\pi\left(\delta L_u \frac{f_{1}}{c}-\delta L_d \frac{f_{4}}{c}\right), \quad 2\pi\left(\delta L_u \frac{f_{2}}{c}-\delta L_d \frac{f_{3}}{c}\right)\,.
\end{equation}
The difference of the readout phases of the SCIs is then proportional to 
\begin{equation}
2\pi\left(\delta L_u\frac{f_{1}-f_{2}}{c}-\delta L_d\frac{f_{4}-f_{3}}{c}\right)\,.
\end{equation}
Since the wavelengths of the beatnote frequencies 
\begin{equation}
\lambda_{ij}=\frac{c}{f_{i}-f_{j}}
\end{equation} 
is large compared to the amplitude of vibrations $\delta L_u,\delta L_d$, the common-mode noise is strongly suppressed. We demonstrate that by measuring the phase noise using two additional photodetectors, see below. 


\subsection{Secondary phase-lock for $f_0$}

Compared to the frequency $f_m$, the stability requirements
for $f_0$ are much less stringent, as it drops out of the final
result. Nevertheless, the phase fluctuations should be small compared to
$\pi/(n/2)$, because otherwise they will wash out the fringes of the
individual interferometers ($n/2$ is the order of the Bragg diffraction). However, $f_0$ is the difference of 
frequencies transmitted in separate optical fibers. Since the optical path
length fluctuations of the fibers are essentially uncorrelated, they
fully contribute to the noise in $f_0$; there is no common-mode
rejection as for $f_r$. Therefore, a stabilization is also
necessary for the phase of the $f_0$ component.

Since the phase fluctuations in $f_0$ exceed $\pi$ even on timescales below a second, the phase at the
start of each pulse is unknown, i.e., it may be
anything between $0$ and $2\pi$. For the
truncated Gaussian pulses with $\sigma \sim 15\,\mu$s and a total
duration of $\sim 200\,\mu$s, lock must be acquired in much below
a microsecond so that the phase is unstabilized only over a
negligible fraction of the pulse duration.

Because of the speed requirement, an electro-optic modulator (New Focus) installed before one of the fiber inputs is used as a phase shifter (Fig. \ref{conjugate}). It requires a voltage of about 400\,V for generating a phase shift of $2\pi$. For reasons of causality, the
minimum time required for acquiring lock is 2-3 times the total
delay in the feedback loop. This contains about 40\,ns delay due to
the optical path length and signal cables. It is thus desired to have the same or higher speed in the electronics used for lock. Especially the high-voltage amplifier is usually rather slow. Thus, we built an amplifier capable of generating 400\,V signals with $t_r=50$\,ns rise-time, using electron tubes in the output stage \cite{EOM}. This is about 10 times faster than the hybrid technology high-voltage operational amplifiers that are otherwise used. The PI servo is built using 200\,MHz gain-bandwidth product operational amplifiers and has a negligible contribution to the overall delay. The system acquires lock with a time constant of about
0.25\,$\mu$s. It is switched off in the inter-pulse interval.

\begin{figure}
\centering \epsfig{file=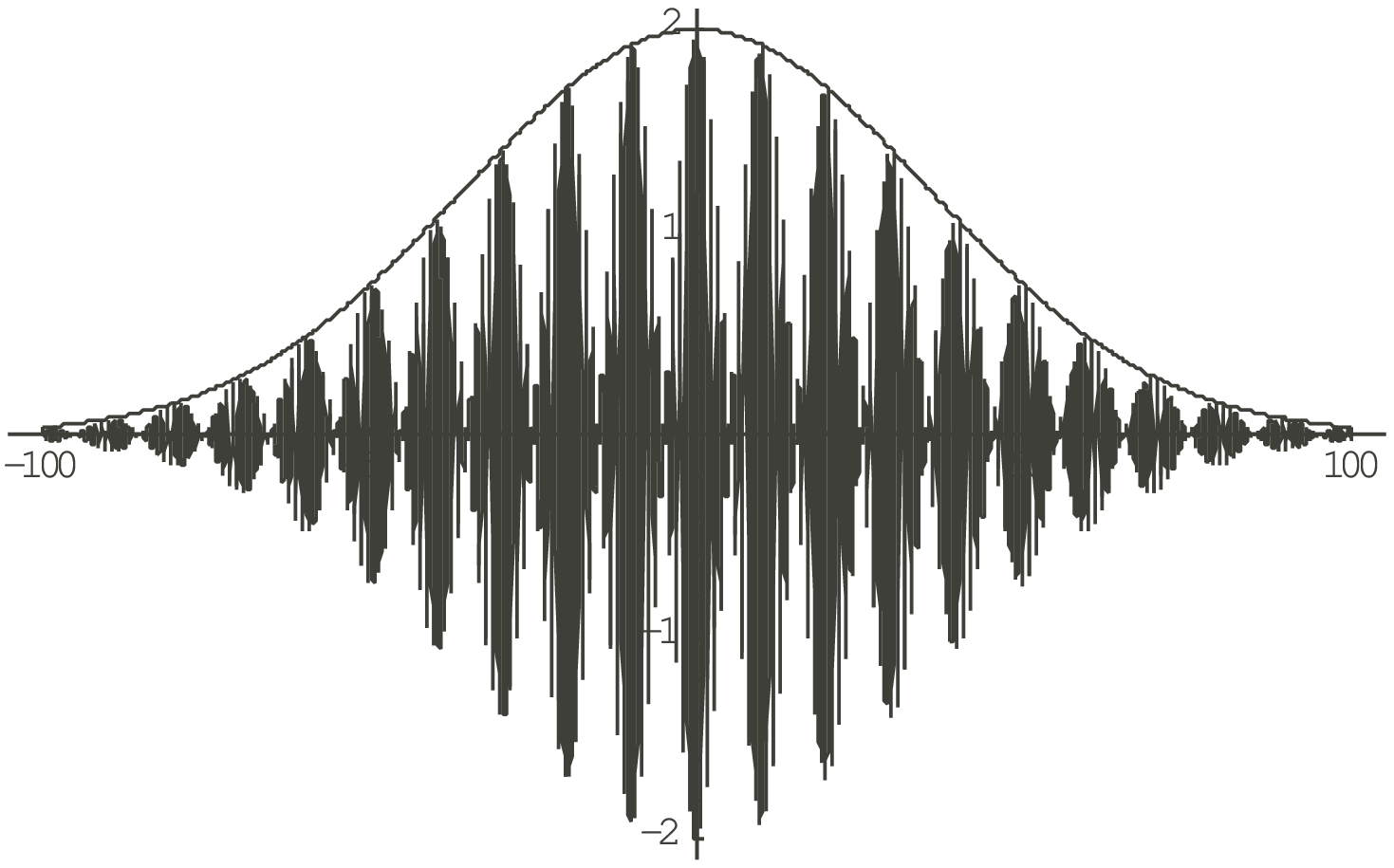,width=0.5\textwidth}
\epsfig{file=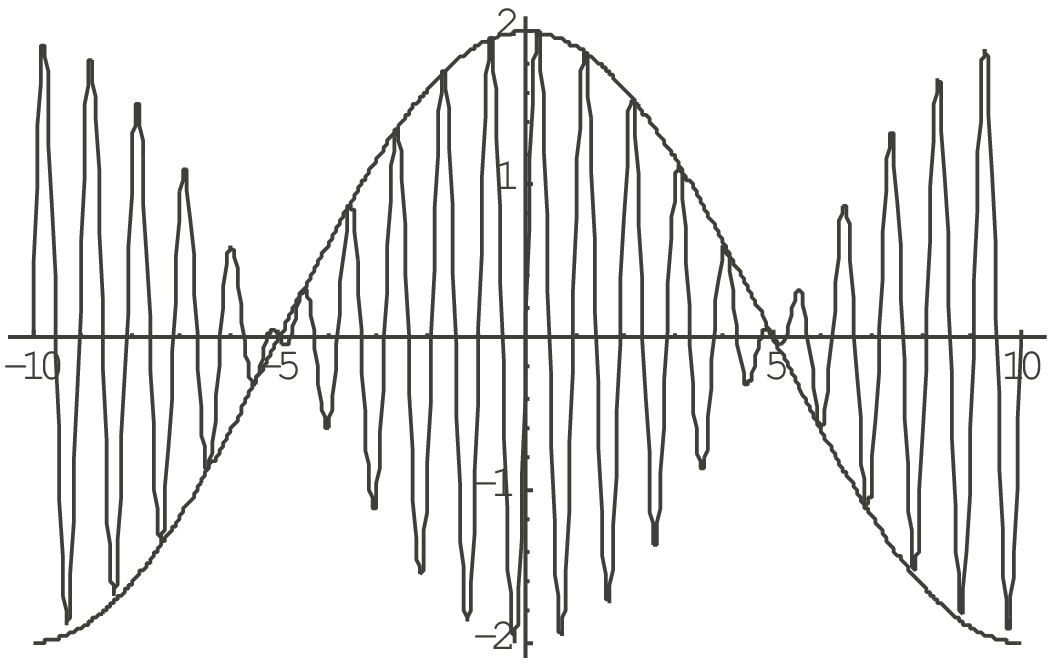,width=0.5\textwidth}
\caption{\label{pulse} Above: Schematic beat signal (not to scale). This signal is obtained from the beat between the two frequency pairs at the fiber output after high-pass filtering to remove a component at $f_m$. Below: Enlarged portion showing an envelope at $f_m$ and an
oscillation at $f_0$. }
\end{figure}

\subsubsection{Signal processing}

The beat signal after the fibers consists of six Fourier components (compare Fig.
\ref{freqgen}): two at $f_m$ that add constructively, one each at
$f_0-f_m$ and $f_0+f_m$, and two components at $f_0$ that cancel
each other if the amplitudes are properly balanced. The $f_m$
component has a frequency $\lesssim 1$\,MHz while the others are above 5\,MHz. Thus, $f_m$ can be
suppressed by a high-order high pass filter. The remaining signal is shown in the time-domain in
Fig. \ref{pulse} (above): It has an overall Gaussian envelope. The enlarged portion shown in Fig.
\ref{pulse} (below) reveals the $f_m$ signal as an envelope of an
oscillation at $f_0$. 

Since the phase of $f_m$ is already
extremely accurately controlled, we are left with stabilizing the
'carrier' at $f_0$. However, at the zeroes of the $f_m$ envelope,
the phase of the $f_0$ signal is switched by 180$^\circ$. This 
reverses the sign of the feedback loop for the $f_0$ phase.
Therefore, the sign has to be electronically reversed
synchronously in order to obtain a stable lock. 
To do that, we synthesize an appropriate reference signal with appropriate polarity
switching by feeding electronic reference signals at $f_0$ and $f_m$ into the two inputs of DBM3 (Fig. \ref{conjugate}). The output signal is converted to a square-wave of constant amplitude by a limiter. 

Because of the high speed of the servo loop, the synchronism between the synthesized signal and the zero-crossings in the optically generated beat signal depicted in Fig. \ref{pulse} must be better than about 200\,ns, or the servo could make cycle slips at each zero-crossing of $f_m$. At $f_m=10$\,kHz, for example, 200\,ns correspond to 2\,mrad phase shift that would be generated by a 1.6\,nm motion of one mirror in the optical setup. It is thus important that the accuracy of the secondary phase lock for $f_m$ is much better than 1\,mrad.

\begin{figure}
\centering
\epsfig{file=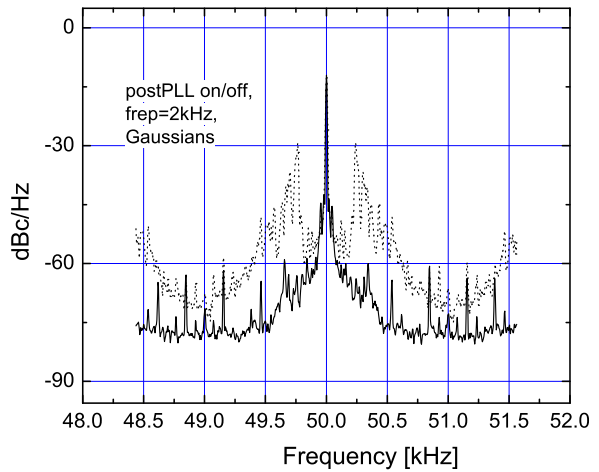,width=0.5\textwidth}
\caption{\label{postPLL} Noise suppression for $f_0$
by secondary phase lock. Dotted graph; secondary phase lock off. Solid graph; secondary phase lock on.}
\end{figure}

The performance of the system is shown in Fig. \ref{postPLL}: In the unstabilized case (upper graph), the phase noise is  strongest at about 200\,Hz offset from the carrier, due to mechanical resonances of mirror
mounts. With stabilization (lower graph), this noise is suppressed by about 30\,dB to a standard
error of $\sqrt{\sigma^2} \sim 0.05$\,rad. This is low
compared to $\pi/(n/2)$ even for $n=10$ photon transitions.

\subsubsection{Vibration cancellation}

As $f_0$ accounts for the Doppler effect due to the free-fall
motion of the atoms and therefore also for the Doppler shift due
to vibrations, such a stabilization will also strongly reduce the
sensitivity of the individual interferometers to vibrations, thus
making an active vibration isolation unnecessary (a simple passive
stage at the bottom of the interferometer is sufficient). This is
because the phase of $f_0$ can be measured after the beams have
passed the interferometers, so that the measurement contains the
vibrationally induced fluctuations. Feedback to the laser phase
can then reduce them. Without the lock, a high-performance vibration isolation would be required fo all optics on top and at the bottom of the fountain \cite{Hensleyvib,Peters,Petersmetrol}.  

\subsection{Performance of the system and conclusions}

With the secondary PLL for $f_m$ (Fig.\ \ref{conjugate}), we measure the residual noise in the differential phase. To make this measurement as realistic as possible, we use another two beat detectors at locations in the optical path that are close to the vacuum chamber. We add their frequencies using a DBM to obtain $2f_m$. We find the noise shown as dashed graph in Fig.\ \ref{PLLphasenoise}. At 1\,Hz offset from the carrier, it has a level of -90\,dBc/Hz. It falls down to about -120\,dBc/Hz at 100\,Hz. At 400\,kHz, it starts to drop again, eventually reaching -138\,dBc/Hz at 1\,MHz. For the interferometers, the laser beams are applied in the form of Gaussian pulses having a width of $\sigma$ and a spacing of $T$. The noise that enters the phase readout can thus be approximated by integrating the noise spectral density from $2/(\pi T)\approx 3$\,Hz to $1/(2\pi\sigma)\approx 10$\,kHz. Thus, we obtain a rms phase uncertainty of $1.6\times10^{-4}$\,rad in $2f_m$, and $8\times 10^{-5}\,$rad in $f_m$. For $n=N=10$ and $T=0.2$\,s in the SCIs, this will contribute 0.15\,ppb of uncertainty in the SCIs' phase, see Eq. \ref{phasenoise}. Laser noise will thus no longer be a limiting factor. 

Usually in atom interferometers, this phase uncertainty is more close to $\sim 0.1-1\,$rad per launch, mainly due to the vibration of mirrors. Thus, a resolution $\sim$1\,ppb could only be reached by integrating over data taken over a period of weeks. Our improvement is due to the combination of the common-mode rejection inherent to SCIs, due to a setup that provides cancellation of path-length fluctuations, and due to the phase-locks described above, which lock precisely the significant phase of $f_m$. 

Note that the phase noise of -90\,dBc/Hz at 1\,Hz corresponds to a relative frequency noise of $10^{-19}$ at 1\,s integration time. For comparison, this level of relative stability is the one reached in gravitational wave detectors like GEO600 and LIGO for time-scales below about 20ms, where shielding from vibrational noise becomes effective. 

\section{Summary and Outlook}

We have described our project to measure $h/m_{\rm Cs}$, the ratio of the Planck constant to the cesium mass to an accuracy of below 0.5 parts per billion. Such a measurement would allow us to determine the fine structure constant $\alpha$ and to test the theory of quantum electrodynamics. After a brief outline of the project, we concentrated on describing the laser system that will provide for the atomic beam splitters and the readout of the atomic phase. The emphasis is on the feedback loops that are used to reduce the laser phase-noise at the location of the atoms. We did not describe the system that will ensure accurate counterpropagation of the beams, as this is described elsewhere \cite{BPM}. Also, we did not describe the operation of the atomic fountain.

The interferometer laser system depicts an effective phase uncertainty of $8 \times 10^{-5}$\,mrad for one single interferometer run, which corresponds to an improvement by a factor of $10^{3-4}$ relative to previous experiments. This provides for an inaccuracy of 0.3\,ppb in a single measurement. This demonstrates the potential of SCIs to increase the resolution in atom interferometers.
 
Besides various sources of noise, many systematic effects will have to be considered. Examples for the present experiments are discussed in \cite{Wicht,Wichtproc,Biraben}. The present experiment has significantly reduced influences of laser phase noise, beam misalignment (by stabilization of the relative propagation direction \cite{BPM}), and the index of refraction of cold atoms studied in \cite{Ketterle} (by going to a relatively large detuning). The characterization and removal of systematic errors is facilitated by low noise, as this allows to run the experiment at full resolution within a short integration time. Improved accuracy by about one order of magnitude can thus hopefully be achieved. 

Measuring $h/m_{\rm Cs}$ will be one of the most important goals of future atom interferometry space missions. The microgravity environment there allows longer interaction times and reduces some systematic effects. Since this will allow to make full use of the phase stability we reached for our SCIs, we believe that there is a strong case for the application of SCIs in such experiments.

\acknowledgments

This work is sponsored in part by grants from the Air Force Office of Scientific Research, the National Science Foundation, and the Multi-University Research Initiative. H.M. whishes to thank the Alexander von Humboldt-Foundation for their support.

\end{document}